\documentstyle[aps,tighten]{revtex}
\begin{document}
\draft
\preprint{UTPT-98-02}
\title{Beyond The Standard Model}
\author{J. W. Moffat}
\address{Department of Physics, University of Toronto,
Toronto, Ontario M5S 1A7, Canada}

\date{\today}
\maketitle

\begin{abstract}%
An overview of unified theory models that extend the standard model is given. A scenario
describing the physics beyond the standard model is developed based on a finite
quantum field theory (FQFT) and the group G=$SO(3,1)\otimes SU(3)\otimes SU(2)\otimes
U(1)$. The field theory is Poincar\'e invariant, gauge invariant, finite and unitary to all orders of
perturbation theory and has a fundamental scale which is chosen to be $\Lambda_F
=1/\sqrt{G_F}\sim 300$ GeV, where $G_F$ is the Fermi coupling constant. The physical Higgs
particle is protected from acquiring a large mass beyond $\sim 1$ TeV, removing the gauge
hierarchy problem associated with the scalar Higgs field. This avoids the need for a composite
Higgs field or supersymmetry. The coupling constants and the fermion masses can be calculated
from a set of low-energy relativistic eigenvalue equations based on truncated Green's functions
and the FQFT, reducing the number of free parameters in the model without a grand unification
scheme. The proton is predicted to be stable. Quantum gravity is perturbatively finite and unitary
to all orders.
\end{abstract}
\vskip 0.2 true in

\pacs{ }

\section{Introduction}
\vskip 0.3 true in
1. { \bf Why Go Beyond the Standard Model?}
\vskip 0.3 true in
The standard model  has been verified to remarkable accuracy\cite{Altarelli} down to
scales of $10^{-15}$ cm, corresponding to energies up to $\simeq 100$ GeV. With the discovery
of the top quark with a mass: $m_t=175.6(5.5)$ GeV all the required fermions in the standard
model are now in place. However, the Higgs particle is yet to be discovered and represents a vital
missing element in the standard model. 

In view of the extraordinary success of the standard model, why are we not satisfied with the
theory? When the Higgs particle is found, we could just declare that particle physics is closed.
The main reason is that there are conceptual difficulties with the standard model, which seem to
indicate that there is new physics beyond it.  Except for certain exceptions,
which will be elaborated upon shortly,  model building beyond the standard model is writ in
theory and the predictions of this model building are often in strong disagreement with
experiments. We shall consider and review the main features of models that go beyond the
standard model.

The main reasons for considering new physics beyond the standard model are:

\begin{enumerate}

\item The gauge hierarchy (or 't Hooft naturalness) problem which besets the Higgs sector.
The standard model cannot naturally explain the relative smallness of the weak scale
of mass, set by the Higgs mechanism at $M_{\rm WS}\sim 250$ GeV.

\item The Dirac naturalness problem: Why is some coupling or mass very small or
even zero when there is no {\it a priori} reason for it to be so? It is expected that all couplings
and interactions which are not otherwise forbidden should be allowed, and that all ratios of
couplings and masses would be ${\cal O}$ (1).

\item Why are there so many arbitrary parameters in the standard model? How can we reduce the
number of unknown parameters?

\item What is the origin of the three fermion generations in the standard model?

\item How can the fermion masses in the standard model be predicted?

\end{enumerate}
\vskip 0.3 true in
2. {\bf The Higgs Sector Hierarchy Problem}
\vskip 0.3 true in
The gauge hierarchy problem\cite{Susskind} is related to the spin $0^+$
scalar field nature of the Higgs particle in the standard model with quadratic mass divergence and
no protective extra symmetry at $m=0$. In standard point particle, local field theory the fermion
masses are logarithmically divergent and there exists a chiral symmetry restoration at
$m=0$.

There have been two main theoretical ways proposed to overcome the hierarchy problem
associated with the Higgs sector:

\begin{enumerate}

\item Higgs compositeness of some kind. The Higgs particle is not elementary but is composed
of  fermions or a condensate, bound by some new strong force, much stronger than the usual
strong interaction.

\item The introduction of supersymmetry.

\end{enumerate}

In the case of (1), many new hadrons are predicted in this kind of scheme in the TeV range.
Several technicolor models have been proposed\cite{Weinberg}.
These models used the breaking of chiral symmetry in massless QCD induced by quark
condensates. For electroweak breaking new heavy technicolor quarks must be postulated to exist
and the QCD scale must be about three orders of magnitude larger. Such a large force postulated
to exist near the electroweak scale $\sim 100$ GeV strongly influences the electroweak precision
tests and is presently in conflict with them\cite{Ellis}, except for artificially fine-tuned models.
\vskip 1.0 true in
3. {\bf Supersymmetric Models}
\vskip 0.3 true in
A plethora of new hadrons is predicted in supersymmetry schemes in the TeV range. The degree
of divergence of the Higgs quadratic self-mass in point particle supersymmetry theories is
reduced from
quadratic to logarithmic avoiding the gauge hierarchy problem. Naturalness demands that the
cutoff be replaced by the splitting between the normal particles and their supersymmetric partners
and that this splitting (times the size of the weak gauge coupling) is of order a few TeV. Then
the masses of most supersymmetric partners would fall within the accessible range of the LHC.
The simplest model of this kind is the minimal standard supersymmetric model 
(MSSM)\cite{Nilles,Dienes}. 

With unbroken supersymmetry the minimal extension of the standard model has fewer
parameters than the standard model itself. However, supersymmetry must be broken which
increases considerably the number of
free parameters. Some symmetries exist in the standard model at the renormalizable level
including, for example, baryon number $B$, lepton number $L$, and symmetries which forbid
flavor-changing neutral currents up to small corrections arising from Yukawa loop couplings.
The MSSM does
not possess these symmetries and, e.g., the proton is predicted to decay with a weak interaction
lifetime and large flavor-changing neutral currents are generated. The proton decay can be
prohibited by introducing new symmetries, e.g. by matter 
parity\cite{Farrar}, although these symmetries are introduced in an {\it ad hoc}
way. After such symmetries are imposed on MSSM, there are still many new, arbitrary
parameters above and beyond the standard model. The flavor-changing neutral current problem
can be fixed by degeneracy, alignment and decoupling\cite{Hall,Babu,Ross}, although these
``cures" are also speculative.

One of the major problems with supersymmetry is that to conform with nature, it must clearly be
broken or hidden. The symmetry should be broken spontaneously, for otherwise all of the
desirable features of symmetries are lost, e.g. renormalizability and unitarity. The spontaneous
symmetry breaking problem turns out to be a severe difficulty of the theory, because the
tree-level supertrace formula:
\begin{equation}
{\rm STR}(M^2)=0,
\end{equation}
is not experimentally tenable, since scalar masses would exist uniformly above and below the
fermion masses. Moreover, the supersymmetry breaking is well above experimental detection,
and there arises the problem of communicating the breaking to the visible part of the standard
model. One
way to address this problem is to use supergravity theory, which is a non-renormalizable theory
containing the spin-$2$ graviton and its spin-$3/2$ partner, the gravitino. This now introduces
gravity into the picture generating a source of flavor physics between the Planck scale and the
weak scales, which can violate the degeneracy or mass universality at the weak
scale\cite{Dienes}.
\vskip 0.3 true in
4. {\bf Grand Unified Theory Models}
\vskip 0.3 true in
The idea of grand unified theories (GUTs)\cite{Salam} appears compelling because it explains:
(i) the quantization of charge, (ii) a relationship between quarks and leptons with $Q_e=-Q_p$,
and (iii) the existence of electromagnetic, weak and strong forces with the coupling constant
relations $g_3 \gg g_2 \gg g_1$ at low energies and $g_1=g_2=g_3$ at high energies. However,
it does not resolve the hierarchy problem, family replication or the origin of fermion masses. The
conventional approach is to assume that the quarks, leptons and the Higgs bosons are elementary
point particles. An alternative approach is to assume that they have some composite structure 
of a constituent called the ``preon"\cite{Pati}. There are two reasons for excluding standard
GUTs, namely, the fact that proton decay searches at the IMB and the Kamiokande 
detectors\cite{Barnett} have failed to detect proton decay, and that precision measurements of
the standard model coupling constants and $\sin^2\theta_W$ at LEP have severely
constrained GUTS without supersymmetry, and excluded them beyond any doubt.

If supersymmetry is invoked then the MSSM model can lead to a meeting of the coupling
constants
with an {\it assumed} supersymmetry threshold $\sim 1$ TeV. The minimal supersymmetric
$SU(5)$ or $SO(10)$ predict a meeting of the coupling constants at a scale $M_U\approx
2\times 10^{16}$ GeV, and for their continuing to stay together beyond this
energy\cite{Dimopolous}. The
predicted value of $[\sin^2\theta_W(m_Z)]_{\rm theory}=0.2325\pm 0.005$ using
$\alpha(m_z)=0.12\pm0.01$ agrees well with the experimentally determined value at LEP:
$[\sin^2\theta_W(m_Z)]_{\rm expt}=0.2316\pm0.0003$\cite{Altarelli}. However, this is a
prediction of only
one number, and it only includes the gauge forces but not the Higgs exchange forces, which still
have arbitrary parameters associated with the masses, the quartic and the Yukawa couplings.
Is this prediction and its agreement with experiment just a coincidence? Of course, if new
experiments do not detect supersymmetric particles at $\sim 1$ TeV, then this whole scenario is
thrown into question. 
\vskip 0.3 true in
5. {\bf Superstring Theory}
\vskip 0.3 true in
A lot of hope has been put into the possibility that the idea of superstrings\cite{Schwarz} will
solve all the problems of unification of the basic forces. The strings are extended objects which
replace the point particles as elementary objects with a size $\sim (M_{\rm Planck})^{-1}
\sim 10^{-33}\,{\rm cm}$ corresponding to the Planck energy $M_{\rm Planck}\sim 10^{19}$
GeV. Superstrings can stabilize the vacuum, avoid large radiative corrections to the quadratic
Higgs mass through sub-Planck scale physics, and avoid tachyonic solutions. They also provide
the possibility of unifying the standard model with gravity and producing a finite quantum
gravity theory. In heterotic string theory, the gauge and gravitational couplings automatically
unify and give a single coupling constant $g_{\rm string}$\cite{Ginsparg}:
\begin{equation}
\label{stringuni}
\frac{8\pi G}{\alpha'}=g_i^2k_i=g^2_{\rm string},
\end{equation}
where $G$ is the Newtonian gravitational constant, $\alpha'$ is the Regge slope, which sets the
mass scale, $g_i$ are the gauge coupling constants, and $k_i$ denote the affine levels.

There has as yet not been any clear cut success of superstring theory at the
phenomenological level, mainly because of the many possible ground state (vacuum) solutions
possible for compactified superstring theory, and because there is no clear understanding of the
necessary breaking of
supersymmetry. Recently, new hopes have been pinned on a theoretical breakthrough from
duality relations\cite{Schwarz2}, which relate perturbative sectors to strong coupling sectors.
There have even been proposals that all string theories and d-brane theories emanate from one
unified theory called M- (or F-) theory.

Difficulties with superstring phenomenology include the predicted superstring unification scale
at one loop level:
\begin{equation}
M_{\rm string}\sim g_{\rm string}\times 5\times 10^{17}\, {\rm GeV}.
\end{equation}
Because the extrapolation of low-energy data indicates that $g_{\rm string}\sim {\cal O}(1)$, we
find $M_{\rm string}\sim 5\times 10 ^{17}$ GeV, which disagrees with the broken GUT scale
of $2\times 10^{16}$ GeV. This leads to a prediction for $\sin^2\theta(m_Z)$ which is in serious
disagreement with the LEP data. Moreover, higher dimensional operators in superstring theory
predict fast proton decay. Speculative methods have to be invented to
overcome this incorrect prediction\cite{Dienes2}. Appeals have to be made to non-perturbative
solutions of string theory to overcome the problem but it is difficult at present to assess whether
this represents any real progress.

As in GUTs and MSSMs superstring theory is severely restricted by the observed stability of the
proton and the light neutrino masses, which are suppressed by many orders of 
magnitude\cite{Faraggi}. These two experimental constraints are hard to satisfy in superstring
theory and although the idea of string unification appears to be aesthetically pleasing and
attractive, the experimental constraints on unification could question the ultimate survival of the
program.
\vskip 0.3 true in
6. {\bf Kaluza-Klein Unified Theory}
\vskip 0.3 true in
Standard Kaluza-Klein unified theories\cite{Kaluza} have fallen into disfavor because of
the Atiyah-Hirzebruch-Witten index theorem\cite{Hirzebruch}. This theorem predicts zero
chirality
number for a Kaluza-Klein compactification scheme where the internal symmetry group arises
from a compact Riemannian geometry. It is difficult to reconcile this result with the necessary
complex nature of fermion representations in, for example, $SO(10)$ models\cite{Wetterich}.
\vskip 0.3 true in
\section{\bf Finite Quantum Field Theory Extension of the Standard Model}
\vskip 0.3 true in
From this brief overview we can draw the conclusion that the unified theory approach to physics
beyond the standard model has not as yet been successful. Of course, new breakthroughs in
superstring theory could solve all of the problems that have arisen to date. But if this does not
come about, then the question arises that maybe this approach is not the correct one to adopt.
Could there exist an alternative
theory beyond the standard model which can resolve such problems as the Higgs sector hierarchy
problem, the Dirac naturalness problem, the origin of fermion masses and reduce the number of
arbitrary parameters in the standard model? In the following, we shall develop such a possible
alternative based on a finite quantum field theory (FQFT) obtained from a nonlocal field
theory\cite{Moffat,Moffat2,Moffat3,Woodard,Kleppe,Hand,Woodard2,Clayton}, the
product gauge group, $G=SO(3,1)\otimes SU(3)\otimes SU(2)\otimes U(1)$, which includes
gravity within the local homogeneous Lorentz group $SO(3,1)$, and a basic relativistic
eigenvalue equation obtained from truncated Green's functions\cite{Moffat4}. This theory can
hope to accomplish the following:

\begin{enumerate}

\item Provide a Poincar\'e invariant and gauge invariant field theory including gravity which is
unitary and finite to all orders of perturbation theory.

\item Remove the hierarchy problem associated with the quadratic Higgs mass.

\item Yield all the known agreement of the standard model with experiment.

\item Predict in terms of a relativistic eigenvalue equation using FQFT, all the
fermion masses and gauge coupling constants from low and intermediate energy data.

\item Predict the stability of the proton.

\end{enumerate}

Any nonlocal field theory effects that arise in FQFT will only occur in quantum loop corrections,
i.e. ( in contrast to string theory\cite{Kleppe}) there will be no nonlocal effects predicted at
the tree graph level, so there will not be any violation of causality in classical electromagnetism
or gravity. 

A fundamental energy scale $\Lambda_F$ is introduced via the FQFT and  we postulate that 
\begin{equation}
\label{basicconst}
\Lambda_F=1/\sqrt{G_F}\sim 300\, {\rm GeV},
\end{equation}
where $G_F$ is the weak interaction Fermi coupling constant. The corresponding length scale is
\begin{equation}
\label{length}
\ell_F\sim 10^{-15}\, {\rm cm}.
\end{equation}
This length scale applies universally in our FQFT including gravity.
Thus there are two fundamental length scales in the theory: $\ell_F$ and $\ell_P$ where 
$\ell_P=1/\sqrt{G}\sim 10^{-33}$ cm is the Planck length. The energy scale (\ref{basicconst})
is not a new parameter introduced into the theory, because it is determined by the known
experimental value of $G_F$. It is natural to
postulate that (\ref{basicconst}) is the fundamental energy scale in the particle sector
$SU(3)\otimes SU(2)\otimes U(1)$, and not the Planck energy scale $\Lambda_P\sim 10^{19}$
GeV associated with the gravitational sector, because beyond the top quark mass $m_t=176$
GeV, the theory predicts that no new particles exist with the possible exception of the Higgs
particle. Moreover, this choice of fundamental energy scale, as we shall see, removes the Higgs
sector hierarchy problem.

In this scenario for physics beyond the standard model, there is no need for supersymmetry to
resolve the Higgs quadratic mass hierarchy problem, so the only particles predicted to exist in the
theory are the experimentally known quarks and leptons, and the as yet undiscovered Higgs
boson. This produces a scheme of the utmost economy as far as the number of required basic
particles is concerned. In contrast the supersymmetric models introduce a large number of 
unknown (and undetected) particles, and superstring theory requires the existence of {\it
infinite} towers of particles.

\section{Finite Quantum Field Theory}

A finite quantum field theory based on a nonlocal interaction Lagrangian has been developed 
which is perturbatively finite, unitary and gauge
invariant\cite{Moffat,Moffat2,Moffat3,Woodard,Kleppe,Hand,Woodard2,Clayton}.
The finiteness draws from
the fact that factors of $\exp[{\cal K}(p^2)/2\Lambda_F^2]$ are attached to propagators which
suppress any ultraviolet divergences in Euclidean momentum space. This makes the propagators
finite, e.g. for the photon  propagator\cite{Moffat}, we have
\begin{equation}
D^c_{\mu\nu}(x-y)
=-\frac{\eta_{\mu\nu}}{(2\pi)^4i}\int\frac{d^4k\Pi(k^2)}{k^2-i\epsilon}\exp[ik\cdot(x-y)].
\end{equation}
The function $\Pi(z)$ satisfies the conditions\cite{Efimov,Moffat}:

\begin{enumerate}

\item $\Pi(z)$ is an entire analytic function of order $\frac{1}{2}\leq\gamma\leq1$,

\item $[\Pi(z)]^*=\Pi(z*)$,

\item $\Pi(x) > 0$ for real x,

\item $\int_0^{\infty} dv\Pi(v) < \infty$.

\end{enumerate}

An important development in FQFT was the discovery that gauge invariance and
unitarity can be restored by adding series of higher interactions\cite{Moffat,Moffat2}. The
resulting theory possesses a nonlocal, nonlinear gauge invariance which agrees with the original
local symmetry on shell but is larger off shell. Quantization is performed in the functional
formalism using an analytic and convergent measure factor which retains invariance
under the new symmetry. An
explicit calculation was made of the measure factor in QED\cite{Moffat2}, and it was obtained
to lowest order in Yang-Mills theory\cite{Kleppe}. Kleppe and Woodard\cite{Woodard2}
obtained an ansatz based on the derived dimensionally regulated result when
$\Lambda_F\rightarrow\infty$ which was conjectured to lead to a general functional measure
factor in FQFT gauge theories. 

A convenient formalism which makes the FQFT construction transparent is based on shadow
fields\cite{Kleppe,Woodard2}. Let us denote by $f_i$ a generic local field and write the standard
local action as
\begin{equation}
S[f]=S_F[f]+S_I[f],
\end{equation}
where $S_F$ and $S_I$ denote the free part and the interaction part
of the action, respectively, and
\begin{equation}
S_F=\frac{1}{2}\int d^4xf_i(x){\cal K}_{ij}f_j(x).
\end{equation}
In a gauge theory $S$ would be the Becchi, Rouet and Stora (BRS) gauge-fixed action including
ghost fields in the invariant action required to fix the gauge.  The kinetic operator ${\cal K}$ is
fixed by defining a Lorentz-invariant distribution operator:
\begin{equation}
{\cal E}\equiv \exp\biggl(\frac{{\cal K}}{2\Lambda_F^2}\biggr)
\end{equation}
and the shadow operator:
\begin{equation}
{\cal O}^{-1}=\frac{{\cal K}}{{\cal E}^2-1},
\end{equation}
where $\Lambda_F$ is a fixed energy scale parameter.

Every field $f_i$ has an auxiliary counterpart field $h_i$, and they are used to form a new action:
\begin{equation}
S[f,h]\equiv S_F[F]-A[h]+S_I[f+h],
\end{equation}
where
\begin{mathletters}
\begin{eqnarray}
F&=&{\cal E}^{-1}f,\\
A[h]&=&\frac{1}{2}\int d^4xf_i{\cal O}^{-1}_{ij}f_j.
\end{eqnarray}
\end{mathletters}

By iterating the equation
\begin{equation}
h_i={\cal O}_{ij}\frac{\delta S_I[f+h]}{\delta h_j}
\end{equation}
the shadow fields can be determined as functionals, and the regulated action is derived from
\begin{equation}
\hat S[f]=S[f,h(f)].
\end{equation}
We recover the original local action when we take the limit $\Lambda_F\rightarrow\infty$ and
$\hat f\rightarrow f, h(f)\rightarrow 0$.

Quantization is performed using the definition
\begin{equation}
\langle 0\vert T^*(O[f])\vert 0\rangle_{\cal E}=\int[Df]\mu[f]({\rm gauge\, fixing})
O[F]\exp(i\hat S[f]).
\end{equation}
On the left-hand side we have the regulated vacuum expectation value of the
$T^*$-ordered product of an arbitrary operator $O[f]$ formed from the local fields $f_i$. The
subscript ${\cal E}$ signifies that a
regulating Lorentz distribution has been used. Moreover, $\mu[f]$ is a measure factor and there
is a gauge fixing factor, both of which are needed to maintain perturbative unitarity in
gauge theories.

The new Feynman rules for QFTF are obtained as follows: The vertices remain unchanged but
every leg of a diagram is connected either to a regularized propagator,
\begin{equation}
\label{regpropagator}
\frac{i{\cal E}^2}{{\cal K}+i\epsilon}
=-I\int^{\infty}_1\frac{d\tau}{\Lambda_F^2}\exp\biggl(\tau
\frac{{\cal K}}{\Lambda^2_F}\biggr),
\end{equation}
or to a shadow propagator,
\begin{equation}
-i{\cal O}=\frac{i(1-{\cal E}^2)}{{\cal K}}=-i\int^1_0\frac{d\tau}{\Lambda_F^2}
\exp\biggl(\tau\frac{{\cal K}}{\Lambda_F^2}\biggr).
\end{equation}
The formalism is set up in Minkowski spacetime and loop integrals are formally
defined in Euclidean space by performing a Wick rotation. This fascilitates the analytic
continuation; the whole formalism could from the outset be developed in Euclidean space.

In FQFT renormalization is carried out as in any other field theory. The bare parameters are
calculated from the renormalized ones and $\Lambda_F$, such that the limit
$\Lambda_F\rightarrow\infty$ is finite for all noncoincident Green's functions, and the bare
parameters are those of the local theory. The regularizing interactions {\it are determined by the
local operators.}

The regulating Lorentz distribution function ${\cal E}$ must be chosen to perform an explicit
calculation in perturbation theory. We do not know the unique choice of ${\cal E}$.
It maybe that there exists an equivalence mapping between all the possible
distribution functions ${\cal E}$. However, once a choice for the function is made, then the
theory and the perturbative calculations are uniquely fixed. A standard
choice in early FQFT papers\cite{Moffat,Moffat2} is
\begin{equation}
\label{reg}
{\cal E}_m=\exp\biggl(\frac{\Box-m^2}{2\Lambda_F^2}\biggr),
\end{equation}
so that for spinor operators
\begin{equation}
\Psi(x)={\cal E}^{-1}_m\psi,\quad \bar\Psi(x)={\cal E}^{-1}_m\bar\psi(x).
\end{equation}

An explicit construction for QED\cite{Moffat2} was given using the Cutkosky rules as applied to
FQFT whose propagators have poles only where ${\cal K}=0$ and whose vertices are entire
functions of ${\cal K}$. The regulated action $\hat S[f]$ satisfies these requirements which
guarantees unitarity on the physical space of states. The local action is gauge fixed and then a
regularization is performed on the BRS theory. 

The infinitesimal transformation
\begin{equation}
\delta f_i=T_i(f)
\end{equation}
generates a symmetry of the regulated action $S[f]$, and the infinitesimal transformation 
\begin{equation}
\hat\delta f_i={\cal E}^2_{ij}T_j(f+h[f])
\end{equation}
generates a symmetry of ${\hat S}[f]$. It follows that FQFT regularization preserves all
continuous symmetries including supersymmetry. The quantum theory will preserve symmetries
provided a suitable measure factor can be found such that
\begin{equation}
\hat\delta([Df]\mu[f])=0.
\end{equation}
Moreover, the interaction vertices of the measure factor must be entire functions of the operator
${\cal K}$ and they must not destroy the FQFT finiteness.

In FQFT tree order, Green's functions remain local except for external lines which are unity on
shell. It follows immediately that since on-shell tree amplitues are unchanged by the
regularization, $\hat S[f]$ preserves all symmetries of $S[f]$ on shell. Also all loops
contain at least one regularizing propagator and therefore are ultraviolet finite. Shadow fields are
eliminated at the classical level, for functionally integrating over them would produce
divergences from shadow loops. Since shadow field propagators do not contain any poles there is
no need to quantize the shadow fields. Feynman rules for ${\hat S}[f]$ are as simple as those
for local field theory.

Kleppe and Woodard\cite{Woodard2} have calculated FQFT for $\phi^4$ scalar field theory in
four dimensions and $\phi^3$ in six dimensions to two-loop order, and explicitly shown that
problems such as overlapping divergences can be dealt with correctly. There
are no problems with power counting and the scalar field theory case is not more difficult to
implement than dimensional regularization. Indeed, one obtains the FQFT regulated result in the
limit $\Lambda_F\rightarrow\infty$ by replacing the gamma function of the dimensional
regularization result by an incomplete gamma function, evaluated as a simple combination of
Feynman parameters.

We recall that regularization schemes, such as Pauli-Villars or a conventional cut-off
method, only define a finite quantum field theory at the price of violating gauge invariance.
Dimensional regularization or $\zeta$-function regularization are finite only in complex
fractional or complex dimensional space. In contrast, FQFT is finite in a real space with
$D\geq 4$.

Although FQFT is perturbatively unitary, at fixed loop order the scattering amplitudes appear to
violate
bounds imposed by partial wave unitarity, i.e. the projected partial waves $A_{\ell }(s)$, where
$s$ is the Mandelstam center-of-mass energy squared, grows beyond the partial wave unitarity
limit as $s\rightarrow\infty$. Because our tree graphs are the same as the local theory this occurs
only for the loop graphs. The same problem occurs in string theory for both the tree graphs and
the  loop graphs\cite{Soldate} at high energies, and it has been studied by Muzinich and
Soldate\cite{Muzinich}. These studies suggested that the problem
can be resolved by a resummation of the perturbation series.

Efimov has provided a solution to the fixed perturbative loop order unitarity 
problem\cite{Efimov2}. He considered a system of two scalar particles with mass $m$ and the
following
inequalities for the upper bound on the elastic scattering amplitude $M(s,t)$:
\begin{equation}
\vert M(s,t)\vert < C(t_0)s,\quad (\vert t\vert \geq \vert t_0\vert > 0)
\end{equation}
and for the total cross section 
\begin{equation}
\sigma_{\rm tot} \leq C\vert \frac{d}{dt}\ln {\rm Im}M(s,t)\vert_{t=0},\quad 
(s\rightarrow\infty).
\end{equation}
These bounds were obtained by using the unitarity of the S-matrix on the mass shell and the
natural assumption that the imaginary part of the elastic scattering, ${\rm Im}M(s,t)$, is a
differentiable and convex down function in the neighborhood of $t=0$. The analyticity of the
elastic scattering amplitude in the Martin-Lehmann\cite{Eden} ellipse and the locality of the
theory were not used in the derivation of the bounds. 

\section{The Standard Model as a FQFT Theory}

Let us now consider the FQFT version of the standard model based on the symmetry group
$SU(3)\otimes SU(2)\otimes U(1)$. We shall start by restricting ourselves to the electroweak
interactions based on the Weinberg-Salam model\cite{Weinberg2,Salam2}. The regularized
auxiliary Lagrangian takes the form:
\begin{equation}
\label{Standard model}
{\hat{\cal L}}_{\rm SM}={\hat{\cal L}}_{\rm 0\,SM}+{\hat{\cal L}}_{\rm I\,SM},
\end{equation}
where
\begin{mathletters}
\begin{eqnarray}
{\hat{\cal L}}_{\rm 0\,SM}=i[{\bar \Psi}^L\gamma\cdot\partial\Psi^L
+\bar\Psi^R_l\gamma\cdot\partial\Psi^R_l+\bar\Psi^R_{\nu_l}\gamma
\cdot\partial\Psi^R_{\nu_l}]\nonumber\\
-i[\bar\chi^L{\cal O}^{-1}\chi^L+\bar\chi^R_l{\cal O}^{-1}
\chi^R_l+\bar\chi^R_{\nu_l}{\cal O}^{-1}\chi^R_{\nu_l}]
+\frac{1}{2}\{{\cal W}_{a\mu}{\cal K}^{\mu\nu}_{ab}
{\cal W}_{b\nu}-C_{a\mu}[W]({\cal O}^{-1})^{\mu\nu}_{ab}C_{b\nu}[W]\}\nonumber\\
+\frac{1}{2}\{{\cal B}_\mu{\cal K}^{\mu\nu}{\cal B}_\nu
- D_\mu[B]({\cal O}^{-1})^{\mu\nu}D_\nu[B],\\
{\hat{\cal L}}_{\rm I\,SM}=-g_1[J^{a\mu}(W_{a\mu}+C_{a\mu})
+g_2J_Y^\mu(B_\mu+D_\mu)]
+g_1\epsilon_{abc}[(W_{a\nu,\mu}+C_{a\nu,\mu})(W^\mu_b+C^\mu_b)
(W^\nu_c+C^\nu_c)\nonumber\\
-\frac{1}{4}g_1^2\epsilon_{abc}\epsilon_{cde}(W_{a\mu}+C_{a\mu})(W_{b\nu}
+C_{b\nu})(W^\mu_d+C^\mu_d)(W^\nu_e+C^\nu_e)].
\end{eqnarray}
\end{mathletters}
Here, $\Psi(x)\equiv {\cal E}^{-1}\psi(x)$, and
\begin{mathletters}
\begin{eqnarray}
J^\mu_a&=&\frac{1}{2}[(\bar\psi^L+\bar\chi^L)\gamma^\mu\tau_a(\psi^L+\chi^L)]
\quad (a=1,2,3)\\
J^\mu_Y&=&-\frac{1}{2}[(\bar\psi^L+\bar\chi^L)\gamma^\mu(\psi^L+\chi^L)
+(\bar\psi^R+\bar\chi^R)\gamma^\mu(\psi^R+\chi^R)].
\end{eqnarray}
\end{mathletters}
Moreover,
\begin{equation}
{\cal W}_{a\mu}={\cal E}^{-1}W_{a\mu}
\end{equation}
and
\begin{equation}
{\cal B}_\mu={\cal E}^{-1}B_\mu.
\end{equation}
The $\tau_a$ are the $2\times 2$ Hermitian Pauli matrices
\begin{equation}
\tau_1=\left(\matrix{0&1\cr 1&0\cr}\right),\quad 
\tau_2=\left(\matrix{0&-i\cr i&0\cr}\right),\quad
\tau_3=\left(\matrix{1&0\cr 0&-1\cr}\right).
\end{equation}
$\chi$ denotes the spinor shadow field while $W_{a\mu}$ and $B_\mu$ denote the intermediate
charged vector boson field and the real boson field, respectively, and $C_{a\mu}$ and $D_\mu$
their respective shadow fields. 

The kinetic operators ${\cal K}^{\mu\nu}_{ab}$ and ${\cal K}_{\mu\nu}$ are given by
\begin{mathletters}
\begin{eqnarray}
{\cal K}^{\mu\nu}_{ab}=\delta_{ab}(\Box\eta^{\mu\nu}-\partial^\mu\partial^\nu),\\
{\cal K}^{\mu\nu}=\Box\eta^{\mu\nu}-\partial^\mu\partial^\nu.
\end{eqnarray}
\end{mathletters}

The shadow field ${C_a}^\mu$ can be expanded as\cite{Kleppe}:
\begin{equation}
C_{\mu_a}[W]
={\cal O}_{\mu\nu ab}g\epsilon_{bcd}[W^\nu_cW^\gamma_{d,\gamma}+W_{c\gamma}
W_d^{\gamma,\nu}-2W_{c\gamma}W_d^{\nu,\gamma}]+O(g^2,W^3).
\end{equation}
The extended non-Abelian gauge transformation is
\begin{equation}
{\hat\delta}_\theta W^\mu_a=-\theta^{,\mu}_a
+{\cal E}^{2\mu\nu}_{ab}g_1\epsilon_{bcd}(W_{c\nu}+C_{c\nu}[W])\theta_d.
\end{equation}

The quark doublets can be easily incorporated into the scheme by
having a quark field $\psi_q$ and its associated shadow field $\chi_q$. The $W_{3\mu}$ and
$B_\mu$ are linear combinations of the two fields $A_\mu$ and $Z_\mu$:
\begin{mathletters}
\begin{eqnarray}
W_{3\mu}=\cos\theta_WZ_\mu+\sin\theta_WA_\mu,\\
B_\mu=-\sin\theta_WZ_\mu+\cos\theta_WA_\mu,
\end{eqnarray}
\end{mathletters}%
where the angle $\theta_W$ denotes the Weinberg angle. The electroweak coupling constants
$g_1$ and $g_2$ are related to the electric charge $e$ by the standard equation
\begin{equation}
g_1\sin\theta_W=g_2\cos\theta_W=e
\end{equation}
and we use the normalization $\cos\theta_W=g_1/(g_1^2+g_2^2)^{1/2}$.

A summation over all different kinds of leptons is understood: $l=e,\mu,...,$ and the field
$\psi^L$ denotes a two-component left-handed lepton field:
\begin{equation}
\psi^L=\left(\matrix{\psi^L_{\nu_l}\cr\psi^L_l\cr}\right)
\end{equation}
with $\psi^L=1/2(1-\gamma_5)\psi$ and $\psi^R=1/2(1+\gamma_5)\psi$ and,
correspondingly,
\begin{equation}
\bar\psi^L=(\bar\psi^L_{\nu_l},\bar\psi^L_l).
\end{equation}

We must add to the Lagrangian (\ref{Standard model}) a Higgs-boson sector. The FQFT
regularized Higgs Lagrangian density has the form:
\begin{eqnarray}
{\hat{\cal L}}_{\rm H}
=-\Phi^{\dag}\Box\Phi+\rho^{\dag}{\cal O}^{-1}\rho
+\frac{1}{2}i[g_1\tau_a(W_{a\mu}+C_{a\mu})\nonumber\\
+\tau_a^{\dag}(W_{a\mu}+C_{a\mu})^{\dag}
+2g_2(B_\mu+D_\mu)]\partial^\mu(\phi+\rho)^{\dag}(\phi+\rho)
+\frac{1}{4}\{g_1g_2[\tau_a(W_{a\mu}+C_{a\mu})\nonumber\\
+\tau_a^{\dag}(W_{a\mu}
+C_{a\mu})^{\dag}](B^\mu+D^\mu)
+g_1^2\tau_a^{\dag}\tau_b(W_a^\mu+C_a^\mu)^{\dag}(W_{b\mu}
+C_{b\mu})\nonumber\\
+g_2^2(B_\mu+D_\mu)(B^\mu+D^\mu)\}(\phi+\rho)^{\dag}
(\phi+\rho)\nonumber\\
-\mu^2(\phi+\rho)^{\dag}(\phi+\rho)-\lambda[(\phi+\rho)^{\dag}
(\phi+\rho)]^2\nonumber\\
-g_l[(\bar\psi_l^L+\bar\chi_l^L)(\psi_l^R+\chi_l^R)(\phi+\rho)+(\phi+\rho)^{\dagger}
(\bar\psi^R_l+\bar\chi^R_l)(\psi^L_l+\chi^L_l)]\nonumber\\
-g_{\nu_l}[(\bar\psi^L_l+\bar\chi^L_l)(\psi^R_{\nu_l}+\chi^R_{\nu_l})(\phi+\rho)+
(\phi+\rho)^{\dagger}(\bar\psi^R_{\nu_l}+\bar\chi^R_{\nu_l})
(\psi^L_l+\chi^L_l)]+{\rm h.c}.
\end{eqnarray}
The regularized isospin doublet Higgs field is
\begin{equation}
\Phi={\cal E}^{-1}\phi
\end{equation}
and $\rho$ is the Higgs shadow field.

In the quantized theory, $SU(2)\times U(1)$ will be spontaneously broken by the vacuum
expectation value of the Higgs field:
\begin{equation}
\langle(\phi+\rho)\rangle_0=\left(\matrix{0\cr v/\sqrt{2}\cr}\right),
\end{equation}
where $v=(-\mu^2/\lambda)^{1/2}$ is not invariant under $SU(2)\times U(1)$ gauge
transformations, but is invariant under the $U(1)$ gauge transformations of
electromagnetism, thereby preserving a massless photon.

Quantum chromodynamics (QCD) can be incorporated within FQFT with the $SU(3)$ group of
phase transformations on the quark color fields. The regularized Lagrangian is
\begin{eqnarray}
{\hat{\cal L}}_{\rm QCD}
=\bar\Psi_{q}(i\gamma\cdot\partial+m)\Psi_{q}-\bar\chi_q{\cal O}^{-1}\chi_q
-\frac{1}{2}\{{\cal G}_{a\mu}{\cal K}^{\mu\nu}_{ab}{\cal G}_{b\nu}
-N_{a\mu}[G]({\cal O}^{-1})^{\mu\nu}_{ab}N_{b\nu}[G]\}\nonumber\\
-g_3[(\bar\Psi_q+\bar\chi_q)\gamma^\mu T_a(\Psi_q+\chi_q)]
(G_\mu^a+N_\mu^a)+{\cal L}_I(G+N[G]),
\end{eqnarray}
where $\Psi_q={\cal E}^{-1}\psi_q$ denote the three standard model color quark fields,
$G^a_\mu$ ($a=1,...,8$) and $N^a_\mu$ denote the massless gluon field and its shadow
field, respectively, and ${\cal G}^a_\mu={\cal E}^{-1}G^a_\mu$. The $T_a$ are a set of
linearly independent traceless $3\times 3$ matrices. Also, we have
\begin{equation}
{\cal L}_I(G)=g_3\epsilon_{abc}G_{a\nu,\mu}G^\mu_bG^\nu_c
-\frac{1}{4}g_3^2\epsilon_{abc}\epsilon_{cde}G_{a\mu}G_{b\nu}G^\mu_dG^\nu_e.
\end{equation}
The extended gauge transformation is
\begin{equation}
{\hat\delta}_\theta G^\mu_a=-\theta^{,\mu}_a
+{\cal E}^{2\mu\nu}_{ab}g_3\epsilon_{bcd}
(G_{c\nu}+N_{c\nu}[G])\theta_d.
\end{equation}

Let us consider the gauge-fixing problem for the gluon field $G^\mu_a$. The same method
applies to the non-Abelian vector boson field $W_a^\mu$. The BRS gluon field Lagrangian in
Feynman gauge has the form\cite{Kleppe}
\begin{eqnarray}
{\cal L}_{\rm G\,BRS}
=-\frac{1}{2}G_{a\nu,\mu}G_a^{\nu,\mu}-\bar\eta_a^{,\mu}\eta_{a,\mu}
+g_3\epsilon_{abc}\bar\eta_a^{,\mu}G_{b\mu}\eta_c\nonumber\\
+g_3\epsilon_{abc} G_{a\nu,\mu}G^\mu_bG^\nu_c
-\frac{1}{4}g_3^2\epsilon_{abc}\epsilon_{cde}
G_{a\mu}G_{b\nu}G^{\mu}_dG^{\nu}_e,
\end{eqnarray}
where $\bar\eta$ and $\eta$ are the BRS ghost fields. The Lagrangian is invariant under the
global transformations:
\begin{mathletters}
\begin{eqnarray}
\delta G_{a\alpha}&=&(\eta_{a,\alpha}-g_3\epsilon_{abc}G_{b\alpha}\eta_c)\delta\zeta,\\
\delta\eta_a&=&-\frac{1}{2}\epsilon_{abc}\eta_b\eta_c\delta\zeta,\\
\delta\bar\eta_a&=&-G^\mu_{a,\mu}\delta\zeta,
\end{eqnarray}
\end{mathletters}
where $\zeta$ is a constant anti-commuting c-number.

The gluon and ghost kinetic operators are
\begin{mathletters}
\begin{eqnarray}
{\cal K}^{\alpha\beta}_{ab}&=&\delta_{ab}\eta^{\alpha\beta}\Box,\\
{\cal K}_{ab}&=&\delta_{ab}\Box.
\end{eqnarray}
\end{mathletters}

The BRS interaction Lagrangian is given by
\begin{equation}
{\cal L}_{\rm I\,BRS}
=g_3\epsilon_{abc}(\bar\eta_a^{,\mu}G_{b\mu}\eta_c+G_{a\nu,\mu}G_b^\mu
G_c^\nu-\frac{1}{4}g_3\epsilon_{cde}G_{a\mu}G_{b\nu}G^\mu_dG^\nu_e).
\end{equation}

The regularized BRS gluon action is
\begin{eqnarray}
{\hat S}(G,\bar\eta,\eta)=\int d^4x(-\frac{1}{2}{\cal G}_{a\alpha,\beta}
{\cal G}^{\alpha,\beta}_a
-\frac{1}{2}N_{a\alpha}\bar{\cal O}^{-1}N^\alpha_a\nonumber\\
-{\hat{\bar\eta}}^{,\alpha}_a{\hat\eta}_{a,\alpha}-{\hat\sigma}_a{\bar{\cal
O}}^{-1}\hat\sigma_a)
+{\hat S}_{\rm G\,I}(G+N,\bar\eta+\bar\sigma,\eta+\sigma),
\end{eqnarray}
where $\sigma$ is the shadow field associated with the ghost field $\eta$ and
\begin{mathletters}
\begin{eqnarray}
{\hat{\bar\eta}}&=&{\cal E}^{-1}\bar\eta,\\
\hat\eta&=&{\cal E}^{-1}\eta.
\end{eqnarray}
\end{mathletters}

The functional quantization proceeds from the expression:
\begin{equation}
\langle 0\vert T^*(O(G,\bar\eta,\eta)\vert 0\rangle
\rangle_{\cal E}=\int[DG][D\bar\eta][D\eta]\mu(G,\bar\eta,\eta)
O({\cal G},{\hat{\bar\eta}},\hat\eta) \exp[i\hat S(G,\bar\eta,\eta)].
\end{equation}
The measure obtained to first order by Kleppe and Woodard is given by
\begin{equation}
\ln[\mu(G,\bar\eta,\eta)]=-\frac{1}{2}g_3^2\epsilon_{acd}\epsilon_{bcd}\int d^4x
G_{a\mu}{\cal M}G^{\mu}_b+{\cal O}(g^3),
\end{equation}
where
\begin{equation}
{\cal M}=\frac{1}{16\pi^2}\int^1_0 d\tau\frac{\Lambda_F^2}{(\tau+1)^2}
\exp\biggl(\frac{\tau}{\tau+1}\frac{\Box}{\Lambda_F^2}\biggr)\biggl(\frac{2}{\tau+1}
+6\frac{\tau}{\tau+1}-3\biggr).
\end{equation}

\section{Finite Quantum Gravity Theory}

We shall now formulate GR as a FQFT. This problem has been considered
previously in\cite{Moffat,Moffat2} but in the following we will regularize the GR equations
using the shadow field formalism. 
The quantum gravity perturbative theory will be locally gauge invariant under
$SO(3,1)$ transformations, and unitary and finite to all orders. At the classical tree graph level
all unphysical polarization states are decoupled and nonlocal effects will only occur in graviton
and graviton-matter loop graphs. 

The Lagrangian density for gravity will be taken as the partially integrated Hilbert form to
remove second derivatives\cite{Papapetrou}:
\begin{equation}
{\cal L}_E={\kappa}^{-2}{\bf g}^{\mu\nu}\biggl(\Gamma^\alpha_{\mu\beta}
\Gamma^\beta_{\nu\alpha}-\Gamma^\alpha_{\mu\nu}\Gamma^\beta_{\alpha\beta}\biggr),
\end{equation}
where $\kappa=(16\pi G)^{1/2}$, ${\bf g}^{\mu\nu}= (-g)^{1/2}g^{\mu\nu}$, 
$g={\rm Det}(g^{\mu\nu})$, and
\begin{equation}
\Gamma^\alpha_{\mu\nu}=\frac{1}{2}g^{\alpha\lambda}(\partial_\nu
g_{\mu\lambda}+\partial_\mu g_{\nu\lambda}-\partial_\lambda g_{\mu\nu}).
\end{equation}

The symmetrical pseudotensor density is
\begin{equation}
{\bf t}^{\mu\nu}=\theta^{\mu\nu}+\frac{1}{2}\partial_\rho({\bf t}^{\mu\nu,\rho}
+{\bf t}^{\rho\mu,\nu}+{\bf t}^{\rho\nu,\mu}),
\end{equation}
where
\begin{equation}
{\theta^\mu}_\nu=\frac{\partial{\cal L}_E}{\partial{\bf g}^{\alpha\beta}/\partial x^\mu}
\partial{\bf  g}^{\alpha\beta}/\partial x^\nu-{\delta^\mu}_\nu{\cal L}_E.
\end{equation}
The total energy-momentum tensor is
\begin{equation}
\Theta^{\mu\nu}={\bf T}^{\mu\nu}+{\bf t}^{\mu\nu}.
\end{equation}
By using the relation
\begin{equation}
\partial_\lambda {{\bf R}^{\lambda\mu}}_\nu=\frac{1}{2}
({{\bf T}^\mu}_\nu+{\theta^\mu}_\nu),
\end{equation}
where
\begin{equation}
{{\bf R}^{\lambda\mu}}_\nu=\frac{{\partial L}_E}{\partial{\bf g}^{\nu\rho}/\partial
x^\lambda}
{\bf g}^{\mu\rho}-\frac{1}{2}{\delta^\mu}_\nu\frac{\partial{\cal L}_E}
{\partial {\bf g}^{\rho\sigma}/\partial x^\lambda}{\bf g}^{\rho\sigma},
\end{equation}
we also have
\begin{equation}
{\bf t}^{\mu\nu,\rho}=-2({{\bf R}^{\rho\mu}}_\lambda\eta^{\nu\lambda}
-{{\bf R}^{\rho\nu}}_\lambda\eta^{\mu\lambda}).
\end{equation}

Let us assume the de Donder coordinate condition\cite{Donder}:
\begin{equation}
\partial_\nu{\bf g}^{\mu\nu}=0.
\end{equation}
The field equations of gravitation become
\begin{equation}
\Box{\bf g}^{\mu\nu}=\kappa^2\Theta^{\mu\nu}
\end{equation}
and the conservation equations give
\begin{equation}
\partial_\nu\Theta^{\mu\nu}=0.
\end{equation}

The Lagrangian ${\cal L}_E$ now becomes
\begin{equation}
{\cal L}_E=\frac{1}{4\kappa^2}{\bf g}^{\mu\nu}g^{\alpha\lambda}
g^{\beta\rho}(2\partial_\beta g_{\mu\lambda}\partial_\alpha g_{\nu\rho}-2\partial_\alpha
g_{\mu\nu}\partial_\beta g_{\rho\lambda}-\partial_\mu g_{\beta\lambda}\partial_\nu
g_{\alpha\rho}+\partial_\mu g_{\beta\rho}\partial_\nu g_{\alpha\lambda}).
\end{equation}
The de Donder harmonic gauge condition can be implemented by adding the noncovariant term:
\begin{equation}
{\cal L}^\prime=\frac{1}{2\kappa^2}\eta_{\mu\nu}
\partial_\alpha{\bf g}^{\mu\alpha}\partial_\beta{\bf g}^{\nu\beta}.
\end{equation}
The total Lagrangian density is
\begin{equation}
{\cal L}_G={\cal L}^\prime+{\cal L}_E.
\end{equation}

We can expand the ${\bf g}^{\mu\nu}$ about any background field, but for simplicity we shall
expand about Minkowski flat spacetime:
\begin{equation}
{\bf g}^{\mu\nu}=\eta^{\mu\nu}+\kappa\gamma^{\mu\nu}+O(\kappa^2).
\end{equation}
Here, $\gamma^{\mu\nu}$ is given by
\begin{equation}
\gamma^{\mu\nu}(x)=(2\pi)^{-3/2}\int\frac{d^3{\bf p}}{(2\omega)^{1/2}}
[a^{\mu\nu}({\bf p})\exp(-ip\cdot x)+a^{\mu\nu*}({\bf p})\exp(ip\cdot x)].
\end{equation}
We have chosen ${\bf g}^{\mu\nu}$ as the interpolating field, so the regularized free part of the
gravitational Lagrangian density takes the form:
\begin{eqnarray}
{\hat{\cal L}}^0_{\rm G}=-\frac{1}{4}[2\eta_{\alpha\beta}
{\hat\gamma}^{\beta\mu}{\cal K}_{\mu\nu}{\hat\gamma}^{\alpha\nu}-\eta^{\mu\nu}
\eta_{\rho\alpha}\eta_{\lambda\beta}{\hat\gamma}^{\lambda\alpha}
{\cal K}_{\mu\nu}{\hat\gamma}^{\rho\beta}\nonumber\\
+\frac{1}{2}\eta^{\mu\nu}\eta_{\lambda\rho}\eta_{\alpha\beta}{\hat\gamma}^{\lambda\rho}
{\cal K}_{\mu\nu}{\hat\gamma}^{\alpha\beta}
+\frac{1}{2}\eta_{\mu\nu}{\hat\gamma}^{\mu\alpha}{\cal K}_{\alpha\beta}
{\hat\gamma}^{\nu\beta},\nonumber\\
-2\eta_{\alpha\beta}s^{\beta\mu}[\gamma]
({\cal O}^{-1})_{\mu\nu}s^{\alpha\nu}[\gamma]
+\eta^{\mu\nu}\eta_{\rho\alpha}\eta_{\lambda\beta}
s^{\lambda\alpha}[\gamma]({\cal O}^{-1})_{\mu\nu}s^{\rho\beta}[\gamma]\nonumber\\
-\frac{1}{2}\eta^{\mu\nu}\eta_{\lambda\rho}\eta_{\alpha\beta}s^{\lambda\rho}[\gamma]
({\cal O}^{-1})_{\mu\nu}s^{\alpha\beta}[\gamma]-\frac{1}{2}\eta_{\mu\nu}
s^{\mu\alpha}[\gamma]({\cal O}^{-1})_{\alpha\beta}s^{\nu\beta}[\gamma]],
\end{eqnarray}
where
\begin{equation}
{\hat\gamma}^{\mu\nu}={\cal E}^{-1}\gamma^{\mu\nu}
\end{equation}
and 
\begin{equation}
{\cal K}_{\mu\nu}=\partial_\mu\partial_\nu. 
\end{equation}
Moreover, $s_{\mu\nu}$ is the shadow field associated with the graviton field
$\gamma_{\mu\nu}$.

The interaction Lagrangian is given by
\begin{eqnarray}
{\hat{\cal L}}^I_{\rm G}
=\frac{\kappa}{4}[\eta_{\rho\alpha}\eta_{\lambda\beta}(\gamma^{\mu\nu}
+s^{\mu\nu})(\partial_\nu(\gamma^{\lambda\alpha}+s^{\lambda\alpha})
\partial_\mu(\gamma^{\rho\beta}+s^{\rho\beta})\nonumber\\
-\frac{1}{2}\eta_{\lambda\rho}\eta_{\alpha\beta}
(\gamma^{\mu\nu}+s^{\mu\nu})\nonumber\\
\partial_\mu(\gamma^{\lambda\rho}
+s^{\lambda\rho})\partial_\nu(\gamma^{\alpha\beta}+s^{\alpha\beta})
+2\eta_{\alpha\lambda}\eta_{\beta\rho}(\gamma^{\alpha\beta}
+s^{\alpha\beta})\nonumber\\
\partial_\nu(\gamma^{\lambda\mu}+s^{\lambda\mu})
\partial_\mu(\gamma^{\rho\nu}
+s^{\rho\nu})+\eta^{\mu\nu}
\eta_{\alpha\beta}\eta_{\sigma\lambda}\eta_{\eta\rho}
(\gamma^{\lambda\rho}
+s^{\lambda\rho})\nonumber\\
\partial_\mu(\gamma^{\alpha\beta}
+s^{\alpha\beta})\partial_\nu(\gamma^{\sigma\eta}
+s^{\sigma\eta})\nonumber\\
-2\eta^{\mu\nu}\eta_{\rho\eta}\eta_{\alpha\lambda}\eta_{\beta\sigma}
(\gamma^{\alpha\beta}+s^{\alpha\beta})
\partial_\nu(\gamma^{\lambda\eta}+s^{\lambda\eta})\partial_\mu
(\gamma^{\sigma\rho}+s^{\sigma\rho})]+O(\kappa^2).
\end{eqnarray}

The graviton regularized propagator is given by
\begin{equation}
D^{\rm grav}_{\mu\nu\lambda\rho}
=(\eta_{\mu\lambda}\eta_{\nu\rho}+\eta_{\mu\rho}\eta_{\nu\lambda}
-\eta_{\mu\nu}\eta_{\lambda\rho})\frac{-i}{(2\pi)^4}
\int d^4k\frac{{\cal E}^2(k^2)}{k^2-i\epsilon}\exp[ik\cdot(x-y)],
\end{equation}
while the shadow propagator is
\begin{equation}
D^{\rm shad}_{\mu\nu\lambda\rho}
=(\eta_{\mu\lambda}\eta_{\nu\rho}+\eta_{\mu\rho}\eta_{\nu\lambda}
-\eta_{\mu\nu}\eta_{\lambda\rho})\frac{-i}{(2\pi)^4}
\int d^4k\frac{[1-{\cal E}^2(k^2)]}{k^2-i\epsilon}\exp[ik\cdot(x-y)].
\end{equation}
In momentum space we have
\begin{mathletters}
\begin{eqnarray}
\frac{-i{\cal E}^2(k^2)}{k^2-i\epsilon}=-i\int^{\infty}_1\frac{d\tau}{\Lambda^2_F}
\exp\biggl(-\tau\frac{k^2}{\Lambda^2_F}\biggr),\\
\frac{i({\cal E}^2(k^2)-1)}{k^2-i\epsilon}=-i\int_0^1\frac{d\tau}{\Lambda^2_F}
\exp\biggl(-\tau\frac{k^2}{\Lambda^2_F}\biggr).
\end{eqnarray}
\end{mathletters}

The regularized Lagrangian will be invariant under nonlinear and field dependent representation
operator transformations much in the same way as for the non-Abelian gauge theory. The
quantization is carried out in the functional formalism by finding a measure factor
$\mu[\kappa\gamma]$ to make
$[D\gamma]$ invariant under the classical symmetry. If we have matter fields, then we must add
measure factors for them as well. To ensure a correct gauge fixing scheme, we express the
Lagrangian in the BRS invariant formalism. The algebra of gauge symmetries is not
expected to
close off-shell, so one needs to introduce higher ghost terms (beyond the normal
ghost terms) into both the action and the BRS transformation. The BRS Lagrangian will be
regularized directly to ensure that all the corrections to the measure factor are included.

\section{Quantum Nonlocal Behavior in FQFT}

In FQFT, it can be
argued that the extended objects that replace point particles (the latter are obtained in the limit
$\Lambda_F\rightarrow\infty$) cannot be probed because of a Heisenberg uncertainty type of
argument.  The FQFT nonlocality {\it only occurs at the quantum loop
level}, so there is no noncausal classical behavior. In FQFT the strength of a signal propagated
over an invariant interval $l^2$ outside the
light cone would be suppressed by a factor $\exp(-l^2\Lambda_F^2)$. 

Nonlocal theories can possess non-perturbative instabilities. These
instabilities arise because of extra canonical degrees of freedom associated with higher time
derivatives. If a Lagrangian contains up to $N$ time derivatives, then the associated Hamiltonian
is linear in $N-1$ of the corresponding canonical variables and extra canonical degrees of
freedom will be generated by the higher time derivatives. The nonlocal theory can be viewed as
the limit $N\rightarrow\infty$ of an Nth derivative Lagrangian. Unless the dependence on the
extra solutions is arbitrarily choppy in the limit, then the higher derivative limit will produce
instabilities\cite{Eliezer}. The condition for the smoothness of the extra solutions is that no
invertible field
redefinition exists which maps the nonlocal field equations into the local ones. String theory does
satisfy this smoothness condition as can be seen by inspection of the S-matrix tree graphs. In
FQFT the tree amplitudes agree with those of the local theory, so the smoothness condition is not
obeyed.

It was proved by Kleppe and Woodard\cite{Kleppe} that the solutions of the nonlocal field
equations are in one-to-one correspondence with those of the original local theory.
The relation for a generic field $v_i$ is
\begin{equation}
v_i^{\rm nonlocal}={\cal E}^2_{ij}v^{\rm local}_j.
\end{equation}
Also the actions satisfy
\begin{equation}
S[v]={\hat S}[{\cal E}^2v].
\end{equation}
Thus, there are no extra classical solutions. The solutions of the regularized nonlocal
Euler-Lagrange equations are in one-to-one correspondence with those of the local action. It
follows {\it that the regularized nonlocal FQFT is free of higher derivative solutions, so FQFT can
be a stable theory.}

Since only the quantum loop graphs in the nolocal FQFT differ from the local field theory,
then FQFT can be viewed as a non-canonical quantization of fields which obey the local
equations of motion. Provided the functional quantization in FQFT is successful, then the theory
does maintain perturbative unitarity. 

\section{Resolution of the Higgs Hierarchy Problem in FQFT}

Let us consider the origin of the `naturalness' problem associated with the Higgs sector.  Consider
the scalar field Lagrangian:
\begin{equation}
{\cal L}_{\phi}=-(\partial_\mu\phi)^\dagger(\partial^\mu\phi)-\mu^2\phi^\dagger\phi
-\lambda(\phi^\dagger\phi)^2,
\end{equation}
where $\phi$ is an $SU(2)$ doublet of complex scalar fields, which is invariant under local
$SU(2)$ gauge transformations. We replace $\partial_\mu$ by the covariant derivative
\begin{equation}
D_\mu=\partial_\mu+ig\frac{\tau_a}{2}W^a_\mu,
\end{equation}
where $W^a_\mu$ are three gauge fields with $a=1,2,3$. Under an infinitesimal gauge
transformation
\begin{equation}
\phi\rightarrow\phi'=(1+i\alpha^a\tau_a/2)\phi,
\end{equation}
the $W^a_\mu$ gauge fields transform as 
\begin{equation}
W^a_\mu\rightarrow
W^a_\mu-\frac{1}{g}\partial_\mu\alpha^a-\epsilon_{abc}\alpha_bW^c_\mu.
\end{equation}
The gauge invariant Lagrangian is
\begin{equation}
{\cal L}_{\phi\,{\rm INV}}
=-\biggl(\partial_\mu\phi+ig\frac{1}{2}\tau_aW^a_\mu\phi\biggr)^\dagger
\biggl(\partial^\mu\phi+ig\frac{1}{2}\tau_a
W^{a\mu}\phi\biggr)-V(\phi)-\frac{1}{4}W^a_{\mu\nu}W_a^{\mu\nu},
\end{equation}
where
\begin{equation}
V(\phi)=\mu^2\phi^\dagger\phi+\lambda(\phi^\dagger\phi)^2,
\end{equation}
and
\begin{equation}
W^a_{\mu\nu}=\partial_\mu W^a_\nu-\partial_\nu W^a_\mu-g\epsilon_{abc}W^b_\mu
W^c_\nu.
\end{equation}
Spontaneous symmetric breaking of $SU(2)$ occurs for $\mu^2 < 0$ and $\lambda > 0$. The
minimum of the potential $V$ occurs when
\begin{equation}
\phi^\dagger\phi=\frac{1}{2}(\phi_1^2+\phi_2^2+\phi_3^2+\phi_4^2)
=-\frac{\mu^2}{2\lambda}.
\end{equation}
We choose
\begin{equation}
\phi_1=\phi_2=\phi_4=0,\quad\phi_3^2=-\frac{\mu^2}{\lambda}\equiv v^2,
\end{equation}
which breaks the $SU(2)$ symmetry spontaneously.  Expanding about this particular vacuum
state we have
\begin{equation}
\langle\phi\rangle_0=\frac{1}{\sqrt{2}}\left(\matrix{0\cr v\cr}\right). 
\end{equation}
The quantum fluctuations about the vacuum $\langle\phi\rangle_0$ can be parametrized by four
real fields $\theta_1,\theta_2,\theta_3,h$, using
\begin{equation}
\phi=\exp(i\tau^a\theta_a/v)\left(\matrix{0\cr\frac{v+h}{\sqrt{2}}\cr}\right).
\end{equation}

Gauging the three massless Goldstone bosons $\theta_a$, we obtain the Lagrangian containing
only physical fields:
\begin{equation}
{\cal L}_{\rm PHYS}
=\frac{1}{2}\partial_\mu h\partial^\mu h+\frac{1}{2}m_H^2h^2-\lambda vh^3
-\frac{1}{4}\lambda h^4+(\rm higher\, order\, terms\, in\, h\, and\, W).
\end{equation}
The field $h$ is the physical neutral scalar Higgs field and $m_H^2=2\lambda v^2$ is the
quadratic Higgs mass. 

Since the Higgs particle is a scalar particle, its mass has radiative contributions which
to one-loop are quadratically divergent. Writing $m_H^2=m_{0H}^2+\delta m_H^2$, where
$m_{0H}$ is the bare Higgs mass and $\delta m_H$ is the Higgs self-energy renormalization
constant, we get for the one loop Feynman graph:
\begin{equation}
\delta m_H^2\sim \frac{\lambda}{32\pi^2}\Lambda_C^2,
\end{equation}
where $\Lambda_C$ is a cutoff parameter. If we want to understand the nature of the Higgs mass
we must require that

\begin{equation}
\delta m_H^2 \leq O(m_H^2),
\end{equation}
i.e. the quadratic divergence should be cut off at the mass scale of the order of the physical Higgs
mass. Since $m_H\simeq \sqrt{2\lambda}v$ and $v=246$ GeV from the electroweak theory,
then in order to keep perturbation theory valid, we must demand that $10\,{\rm GeV} \leq m_H
\leq 350\,{\rm GeV}$ and we need 
\begin{equation}
\label{Higgscut}
\Lambda_C =\Lambda_{\rm Higgs}\leq 1\, {\rm TeV}, 
\end{equation}
where the lower bound on $m_H$ comes from the avoidance of washing out the spontaneous
symmetry breaking of the vacuum.

Nothing in the standard model can tell us why (\ref{Higgscut}) should be true, so we must go
beyond the standard model to solve the problem. $\Lambda_C$ is an arbitrary parameter in point
particle field theory with no physical interpretation. Since all particles interact through gravity,
then ultimately we should expect to include gravity in the standard model, so we expect that
$\Lambda_{\rm Planck}\sim 10^{19}$ GeV should be the natural cutoff. Then we have using
(\ref{Higgscut}) and $\lambda\sim 1$:
\begin{equation}
\frac{\delta m^2(\Lambda_{\rm Higgs})}{\delta m^2(\Lambda_{\rm Planck})}
\approx \frac{\Lambda^2_{\rm Higgs}}{\Lambda^2_{\rm Planck}}\approx 10^{-34},
\end{equation}
which represents an intolerable fine-tuning of parameters. This `naturalness' or hierarchy problem
is one of the most serious defects of the standard model. 

As we discussed earlier, there have been two strategies proposed as ways out of the hierarchy
problem. The Higgs is taken to be composite at a scale $\Lambda_C\simeq 1$ TeV, thereby
providing a natural cutoff in the quadratically divergent Higgs loops. One such scenario is the 
`technicolor' model, but it cannot be reconciled with the accurate standard model data, nor with
the smallness of 
fermion masses and the flavor-changing neutral current interactions. The other strategy is to
postulate supersymmetry, so that the opposite signs of the boson and fermion lines cancel by
means of the non-renormalization theorem. However, supersymmetry is badly broken at lower
energies, so we require that
\begin{equation}
\delta m^2\sim \frac{\lambda}{32\pi^2}\vert\Lambda^2_{C\,{\rm bosons}}
-\Lambda^2_{C\,{\rm fermions}}\vert\leq 1\,{\rm TeV}^2,
\end{equation}
or, in effect
\begin{equation}
\vert m_B-m_F\vert \leq 1\, {\rm TeV}.
\end{equation}
This physical requirement leads to the prediction that the supersymmetric partners of known
particles  should have a threshold $\leq1$ TeV. 

There is no truly convincing experimental result so far that verifies that supersymmetry might be
relevant to
nature. If no supersymmetric particles are discovered at the LHC, then we must seriously begin to
doubt that supersymmetry plays a fundamental role in nature at least at the observable
phenomenological level. The MSSM does not significantly reduce the number of parameters in
the standard model, and as we have observed, there are some serious
phenomenological difficulties with the model which are not easily eradicated. The MSSM model
and superstring theory also introduce a huge
host of new particles which seriously complicates an already complicated situation in the
`particle zoo'.

 A third possible strategy is to introduce a FQFT formalism, and base the whole scheme on the
product group: $SO(3,1)\otimes SU(3)\otimes SU(2)\otimes U(1)$ and realize a field theory
mechanism which will introduce a natural physical scale in the theory $\leq 1$ TeV, which will
protect the Higgs mass from becoming large and unstable.

Let us consider the regularized scalar field FQFT Lagrangian:
\begin{equation}
{\hat{\cal L}}_S=\frac{1}{2}\Phi(\Box-m^2)\Phi
-\frac{1}{2}\rho{\cal O}^{-1}\rho+\frac{1}{2}Z^{-1}\delta m^2(\phi+\rho)^2
-\frac{1}{24}\lambda_0(\phi+\rho)^4,
\end{equation}
where $\phi=Z^{1/2}\phi_R$ is the bare field, $\phi_R$ is the renormalized field, 
$\Phi={\cal E}^{-1}\phi$, $
\rho$ is
the shadow field, $m_0$ is the bare mass, $Z$ is the field strength renormalization constant,
$\delta m^2$ is the mass renormalization constant and $m$ is the physical mass. The
regularizing operator is given by (\ref{reg}), while the shadow kinetic operator is
\begin{equation}
{\cal O}^{-1}=\frac{\Box-m^2}{{\cal E}_m^2-1}.
\end{equation}

The full propagator is
\begin{equation}
-i\Delta_R(p^2)=\frac{-i{\cal E}_m^2}{p^2+m^2-i\epsilon}
=-i\int_1^{\infty}\frac{d\tau}{\Lambda_F^2}\exp\biggl[-\tau\biggl(\frac{p^2+m^2}
{\Lambda_F^2}\biggr)\biggr],
\end{equation}
whereas the shadow propagator is
\begin{equation}
i\Delta_{\rm shadow}
=i\frac{{\cal E}_m^2-1}{p^2+m^2}=-i\int^1_0\frac{d\tau}{\Lambda_F^2}
\exp\biggl[-\tau\biggl(\frac{p^2+m^2}{\Lambda_F^2}\biggr)\biggr].
\end{equation}

Let us define the self-energy $\Sigma(p)$ as a Taylor series expansion around the mass shell
$p^2=-m^2$:
\begin{equation}
\Sigma(p^2)=\Sigma(-m^2)+(p^2+m^2)\frac{\partial\Sigma}{\partial p^2}(-m^2)
+{\tilde \Sigma}(p^2),
\end{equation}
where ${\tilde\Sigma}(p^2)$ is the usual finite part in the point particle limit
$\Lambda_F\rightarrow\infty$.
We have
\begin{equation}
{\tilde\Sigma}(-m^2)=0,
\end{equation}
and 
\begin{equation}
\frac{\partial{\tilde\Sigma}(p^2)}{\partial p^2}(p^2=-m^2)=0.
\end{equation}

The full propagator is related to the self-energy $\Sigma(p^2)$ by
\begin{equation}
-i\Delta_R(p^2)=\frac{-i{\cal E}_m^2[1+{\cal O}\Sigma(p^2)]}{p^2+m^2+\Sigma(p^2)}
=\frac{-iZ}{p^2+m^2+\Sigma_R(p^2)}.
\end{equation}
Here $\Sigma_R(p^2)$ is the renormalized self-energy which can be written as
\begin{equation}
\Sigma_R(p^2)=(p^2+m^2)\biggl[\frac{Z}{{\cal E}_m^2(1+{\cal O}\Sigma)}-1\biggr]
+\frac{Z\Sigma}{{\cal E}_m^2(1+{\cal O}\Sigma)}.
\end{equation}
The 1PI two-point function is given by
\begin{equation}
-i\Gamma_R^{(2)}(p^2)=i[\Delta_R(p^2)]^{-1}=\frac{i[p^2+m^2+\Sigma(p^2)]}
{{\cal E}_m^2[1+{\cal O}\Sigma(p^2)]}.
\end{equation}
Since ${\cal E}_m\rightarrow 1$ and ${\cal O}\rightarrow 0$ as $\Lambda_F\rightarrow\infty$,
then in this limit
\begin{equation}
-i\Gamma_R^{(2)}(p^2)=i[p^2+m^2+\Sigma(p^2)],
\end{equation}
which is the standard point particle result.

The mass renormalization is determined by the propagator pole at $p^2=-m^2$ and we have
\begin{equation}
\Sigma_R(-m^2)=0.
\end{equation}
Also, we have the condition 
\begin{equation}
\frac{\partial\Sigma_R(p^2)}{\partial p^2}(p^2=-m^2)=0.
\end{equation}

The renormalized coupling constant is defined by the four-point function
$\Gamma^{(4)}_R(p_1,p_2,p_3,p_4)$ at the point $p_i=0$:
\begin{equation}
\Gamma_R^{(4)}(0,0,0,0)=\lambda.
\end{equation}
The bare coupling constant $\lambda_0$ is determined by
\begin{equation}
Z^2\lambda_0=\lambda+\delta\lambda(\lambda,m^2,\Lambda^2_F).
\end{equation}
Moreover,
\begin{mathletters}
\begin{eqnarray}
Z&=&1+\delta Z(\lambda,m^2,\Lambda_F^2),\\
Zm_0^2&=&Zm^2-\delta m^2(\lambda,m^2,\Lambda^2_F).
\end{eqnarray}
\end{mathletters}

A calculation of the scalar field mass renormalization gives\cite{Woodard2}:
\begin{equation}
\delta m^2=\frac{\lambda}{32\pi^2}m^2
\Gamma\biggl(-1,\frac{m^2}{\Lambda_F^2}\biggr)+O(\lambda^2),
\end{equation}
where $\Gamma(n,z)$ is the incomplete gamma function:
\begin{equation}
\Gamma(n,z)=\int_z^{\infty}\frac{dt}{t}t^n\exp(-t)=(n-1)\Gamma(n-1,z)+z^{n-1}\exp(-z).
\end{equation}

The renormalization coupling constant $\delta\lambda$ is\cite{Woodard2}:
\begin{equation}
\label{coupl}
\delta\lambda=\frac{3\lambda^2}{16\pi^2}m^2\int_0^{1/2}
dx\Gamma\biggl(0,\frac{1}{1-x}\frac{m^2}{\Lambda_F^2}\biggr)+O(\lambda^3).
\end{equation}

We have
\begin{equation}
\Gamma(-1,z)=-E_i(z)+\frac{1}{z}\exp(-z).
\end{equation}
For small $z$ we obtain the expansion
\begin{equation}
E_i(z)=-\ln(z)-\gamma+z-\frac{z^2}{2\cdot 2!}+\frac{z^3}{3\cdot 3!}-...,
\end{equation}
where $\gamma$ is Euler's constant.  For large positive values of $z$, we have the asymptotic
expansion
\begin{equation}
E_i(z)\sim\exp(-z)\biggl[\frac{1}{z}-\frac{1}{z^2}+\frac{2!}{z^3}-...\biggr].
\end{equation}
Thus, for small $m^2/\Lambda_F^2$ we obtain
\begin{equation}
\label{lambdast}
\delta m^2=\frac{\lambda}{32\pi^2}
\biggl[\Lambda_F^2-m^2\ln\biggl(\frac{\Lambda_F^2}{m^2}\biggr)
-m^2(1-\gamma)+O\biggl(\frac{m^2}{\Lambda_F^2}\biggr)\biggr]+O(\lambda^2),
\end{equation}
while for large values of $m^2/\Lambda_F^2$ we get
\begin{equation}
\delta m^2\sim\frac{\lambda}{32\pi^2}\Lambda^2_F
\exp\biggl(-\frac{m^2}{\Lambda_F^2}\biggr).
\end{equation}
Eq.(\ref{lambdast}) is the standard result for the mass renormalization constant obtained
from a cutoff procedure or dimensional regularization. 

For the renormalization coupling constant we obtain from (\ref{coupl})
for small $m^2/\Lambda_F^2$:
\begin{equation}
\delta\lambda=\frac{3\lambda^2}{32\pi^2}\biggl[\ln\biggl(\frac{\Lambda_F^2}
{m^2}\biggr)+\ln2-1-\gamma+O\biggl(\frac{m^2}{\Lambda_F^2}\biggr)\biggr]+O(\lambda^3),
\end{equation}
which is the familiar result for one loop obtained from dimensional regularization in point
particle field theory. For large positive values of $m^2/\Lambda^2_F$ we get
\begin{equation}
\delta\lambda\sim \frac{3\lambda^2}{32\pi^2}\frac{\Lambda^2_F}{m^2}
\exp\biggl(-\frac{m^2}{\Lambda_F^2}\biggr).
\end{equation}

For $m_H< \Lambda_F$ and for our universal constant $\Lambda_F\sim 300$ GeV, the
physical Higgs mass is protected from becoming too large, while for a heavy Higgs mass,
$m_H >\Lambda_F$, the Higgs self-energy is exponentially damped. Moreover,
the same holds true for the Higgs scalar field coupling constant radiative corrections.  Thus the
scalar Higgs sector is protected from large unstable radiative corrections and FQFT provides a
solution to the naturalness problem, without invoking supersymmetry or Higgs compositeness.
The universal fixed FQFT scale $\Lambda_F$ corresponds to the fundamental length
Eq.(\ref{length}). For a Higgs mass much larger than 1 TeV, the Higgs sector becomes
non-perturbative and we must be concerned about violations of unitarity\cite{Veltman}.

We can calculate the $\beta$ function which is independent of the renormalization conditions.
We have
\begin{equation}
\beta(\lambda)=-\Lambda_F\biggl(\frac{\partial\lambda}{\partial\Lambda_F}
\biggr)_{m,\lambda_0}.
\end{equation}
For small $m^2/\Lambda_F^2$ we obtain to one-loop order\cite{Woodard2}:
\begin{equation}
\beta(\lambda)=\frac{3\lambda^2}{16\pi^2}+O(\lambda^3),
\end{equation}
whereas for large $m^2/\Lambda_F^2$ we find
\begin{equation}
\beta(\lambda)\sim\biggl(\frac{3\lambda^2}{16\pi^2}\biggr)
\exp\biggl(-\frac{m^2}{\Lambda_F^2}\biggr).
\end{equation}
Thus, for the $\beta$ function there is also an exponential damping for large Higgs mass
and for a fixed universal constant $\Lambda_F$.

\section{The Gluon Self-energy}

Kleppe and Woodard have derived the one loop gluon self-energy using the shadow FQFT
formalism\cite{Kleppe}. The one loop vacuum polarization tensor is
\begin{equation}
\Pi^{\alpha\beta}_{ab}=\frac{g_3^2}{16\pi^2}\epsilon_{acd}\epsilon_{bcd}(p^2
\eta^{\alpha\beta}-p^\alpha p^\beta)\Pi(p^2),
\end{equation}
where
\begin{eqnarray}
\Pi(p^2)=2\int_0^{1/2}dx\Gamma(0,xp^2/\Lambda_F^2)[4x(1-x)+1]\nonumber\\
=2\int_0^{1/2}dxE_i(xp^2/\Lambda_F^2)[4x(1-x)+1].
\end{eqnarray}

Because our basic FQFT energy scale is fixed at $\Lambda_F\sim 300$ GeV, we should expect
that QCD and graviton self-energy effects for Euclidean momentum, $p^2 >>
\Lambda^2_F$, are also protected from
becoming large and quantum gravity loop corrections will remain perturbatively weak, even in
the Planck energy range. This is a distinct advantage from a theoretical point of view, for
non-perturbative quantum gravity is not understood on even an elementary qualitative level.

\section{Dynamical Eigenvalue Equation for Fermion Masses}

Our particle physics model is based on the product group $G=SO(3,1)\otimes SU(3)\otimes
SU(2)\otimes U(1)$. There is no attempt at unifying the different particle representations into one
GUT group, so the number of particles predicted to exist is just those of the standard model:
three generations of quarks and leptons, the gluon, the $W$ and $Z $ intermediate boson masses,
the Higgs particle and the graviton.  While we gain in simplicity in having many fewer particles
to contend with than in unified field theory models, we still have to face the large number of
arbitrary parameters associated with the standard model. The origin of fermion masses has
remained a frustrating problem for the standard model, in which they are generated by
Yukawa couplings to the Higgs field with separate coupling constants for each mass. In GUTs
theories the origin of the fermion masses is
pushed to GUT scales or the Planck mass scale, $10^{15}-10^{19}$ GeV, which are many
orders of magnitude above achievable laboratory scales. Superstring theories have not succeeded
in predicting fermion masses and there continues to be a problem as to how the supersymmetry
breaking can coexist naturally with light fermion mass scales.

A way out is to seek a dynamical origin for the fermion masses at lower energy scales $ < 1$
TeV, by deriving a relativistic eigenvalue equation like its nonrelativistic Schr\"odinger equation
counterpart. But there is an important difference between the two equations, for a relativistic
eigenvalue equation would have an infinite number of integral equations which are difficult to
solve. However, it is conceivable that reasonable approximations can be made which could lead
to a satisfactory prediction of the fermion masses. Another difficulty with the approach of
seeking a relativistic eigenvalue equation is that the equation is normally based on a nonlinear
fermion operator interaction which is nonrenormalizable. Past efforts to overcome this problem
have been based on various regularization schemes. Heisenberg and his
collaborators\cite{Heisenberg} overcame the divergence problem by introducing dipole ghost
states into the Hilbert space with an indefinite metric. This meant giving up unitarity at all
energies which is a high price to pay for obtaining a solution to the problem. But with FQFT we
do have a possible solution to the problem, for the field theory is perturbatively finite, unitary and
Poincar\'e invariant to all orders, so we have a self-consistent basis for solving the equations. 

Extensive work has been done in the past to derive the fermion mass spectrum using a
Schwinger-Dyson equation, combined with a Bethe-Salpeter
approximation\cite{Delbourgo}, and a chiral symmetry breaking mechanism.  Fermion masses
are also generated by a Nambu-Jona-Lasinio mechanism\cite{Nambu} using a four-fermion
interaction and a spontaneously broken
vacuum. However, these methods appear to need an arbitrary coupling constant for each fermion
mass, as in the case of the Yukawa couplings in the standard model. We shall approach the
problem of deriving fermion masses using the methods of Heisenberg and his
collaborators\cite{Heisenberg}. However, this approach requires a knowledge of the true
equation of motion for fermion operators with a higher order fermion self-interaction. Heisenberg
based his work on a unified field theory model with a four-fermion interaction. We shall
generalize their
methods so that they are applicable to our extension of the standard model. 
We shall replace the standard model Yukawa couplings to lepton and quark fields by a
four-fermion interaction. It is hoped that the
four-fermion interaction in the equations of motion for the quark or lepton fields is tractable
enough to predict reliable fermion mass spectra in future calculations.

This approach to solving the fermion mass problem was previously considered\cite{Moffat4} but
we shall give an overview of the problem in the context of the FQFT formalism. Let us
consider a local fermion state $\vert\psi\rangle$ which represents a lepton or a quark and the
infinite set of $\tau$-functions:
\begin{equation}
\label{taufunctions}
\tau(x_1x_2...\vert y_1y_2...)=
\langle 0\vert T(\psi(x_1)\psi(x_2)...\bar\psi(y_1)\bar\psi(y_2)...)\vert\psi\rangle.
\end{equation}
The $\tau$-functions describe a covariant representation of the state $\vert\psi\rangle$ and $T$
denotes the time ordered product, which orders operators with larger values of time variable to
the left of the operators with smaller time values. 

A second set of $\tau$-functions can be obtained by setting:
\begin{eqnarray}
\tau(x_1x_2...\vert y_1y_2...)
=\phi(x_1x_2...\vert y_1y_2...)
+\phi(x_1x_2...\vert y_1y_2...)\nonumber\\
(-1)^m\phi(x_2...\vert
y_2...)\langle0\vert T(\psi(x_1)\bar\psi(y_1))\vert 0\rangle
+(-1)^m\phi(x_2...\vert y_1y_3...)\langle 0\vert T(\psi(x_1)\bar\psi(y_2))\vert
0\rangle+...+\nonumber\\
(-1)^m\phi(x_3...\vert y_3...)\langle 0\vert T(\psi(x_1)\bar\psi(y_1))\vert 0\rangle\langle 0\vert
T(\psi(x_2)\bar\psi(y_2))\vert 0\rangle+...,
\end{eqnarray}
where $(-1)^m$ is associated with the number of transpositions necessary to get the operators in
the vacuum expectation values together. If the two-point function is known, then the
$\tau$-functions can be determined.

The formalism invented by Heisenberg and his collaborators\cite{Heisenberg} was based on the
truncated Tamm-Dancoff method (TTD)\cite{Tamm}, in which a truncated set of
$\phi$-functions is defined so that all the $\phi$s of more than $n$ variables are set equal to zero,
yielding an approximation for the physical lepton or quark state $\vert\psi\rangle$. The two-point
function must be approximated to obtain a solvable set of equations. This method was applied to
the nonrelativistic anharmonic oscillator equation of motion, $\ddot q=-\alpha
q^3$\cite{Heisenberg}, and it was found that the eigenvalues do converge with increasing $n$ to
a limiting value which coincides with the exact quantum mechanical solution. It is not known
whether such convergence holds for the more difficult relativistic problem, but we shall assume
in the following that this is true.  

Let us consider the fermion Lagrangian:
\begin{equation}
{\cal L}_{\psi}=\bar\psi\gamma\cdot\partial\psi+{\cal I}(\bar{\psi}^{\cal E},\psi^{\cal E},B),
\end{equation}
where ${\cal I}$ denotes a nonlinear fermion operator self-interaction term, 
\begin{equation}
\psi^{\cal E}={\cal E}\psi,\quad \bar\psi^{\cal E}={\cal E}\bar\psi,
\end{equation}
and $B$ denotes internal symmetry breaking terms. There is no gauge invariance or gauge
fixing connected with the fermion self-energy in the context of the following, so we shall adopt
for convenience the regularization method of ref. (\cite{Moffat2}), although the shadow field
formalism could be equally well applied. The first order dynamical equation of
motion for $\psi$ is 
\begin{equation}
\label{eqsmotion}
\gamma\cdot\partial\psi+{\cal N}({\bar\psi}^{\cal E},\psi^{\cal E},B)=0,
\end{equation}
where
\begin{equation}
{\cal N}({\bar\psi}^{\cal E},\psi^{\cal E},B)=\frac{\partial{\cal I}({\bar\psi}^{\cal E},
\psi^{\cal E},B)}{\partial{\bar\psi}}.
\end{equation}

The local two-point propagator function for the fermions in the Heisenberg representation is
given by
\begin{eqnarray}
S(x-x')=\langle 0\vert T(\psi(x){\bar\psi}(x'))\vert 0\rangle=\nonumber\\
i(2\pi)^{-4}\int d(\kappa^2)\rho(\kappa^2)\int d^4p\frac{\gamma^\nu p_\nu
\exp[-ip\cdot(x-x')]}{p^2+\kappa^2},
\end{eqnarray}
where $\rho(\kappa^2)$ denotes the K\"allen-Lehmann spectral function which contains
delta-functions for the discrete eigenvalues and continuous functions for the continuous spectra. 
The regularized two-point propagator will have the form
\begin{eqnarray}
{\cal S}(x-x')=\langle 0\vert T(\psi^{\cal E}(x){\bar\psi}^{\cal E}(x'))\vert 0\rangle=\nonumber\\
i(2\pi)^{-4}\int d(\kappa^2)\rho(\kappa^2)\int d^4p\frac{\gamma^\nu p_\nu
\Pi(p^2)\exp[-ip\cdot(x-x')]}{p^2+\kappa^2},
\end{eqnarray}
where the function $\Pi(p^2)$ is an entire analytic function which damps out strongly near the
light cone in Euclidean momentum space.

The integrated form of (\ref{eqsmotion}) is
\begin{equation}
\psi(x)=\psi_0(x)-\int d^4x'G(x-x'){\cal N}(x'),
\end{equation}
where
\begin{equation}
G(x-x')=(2\pi)^{-4}\int d^4p\frac{\gamma^\nu p_\nu}{p^2}\exp[-ip\cdot(x-x')]
\end{equation}
is the Green's function belonging to the Weyl equation
\begin{equation}
\label{Weyl}
i\gamma_\nu\biggl(\frac{\partial\psi}{\partial x^\nu}\biggr)=0
\end{equation}
and $\psi_0$ obeys this equation. 

The spinor $\psi$ can be described by left-handed and right-handed Weyl spinors:
\begin{equation}
\label{leftright}
\psi_L=\frac{1}{2}(1-\gamma_5)\psi,\quad \psi_R=\frac{1}{2}(1+\gamma_5)\psi.
\end{equation}
Two simple examples of a four-fermion interaction are
\begin{mathletters}
\begin{eqnarray}
\label{gamma5}
{\cal I}_1({\bar\psi}^{\cal E},\psi^{\cal E})&=&\frac{f^2}{2}[({\bar\psi}^{\cal E}\gamma^\mu
\psi^{\cal E})^2-({\bar\psi}^{\cal E}\gamma^\mu\gamma^5\psi^{\cal E})^2],\\
\label{sigma}
{\cal I}_2({\bar\psi}^{\cal E},\psi^{\cal E})&=&\frac{f^2}{2}({\bar\psi}^{\cal
E}\sigma^{\mu\nu}
\psi^{\cal E})^2,
\end{eqnarray}
\end{mathletters}
where $f^2$ is a coupling constant with the dimensions $[\rm mass]^{-2}$ and $(\quad)^2
=\vert\quad\vert^2$.

Substituting (\ref{leftright}) into (\ref{gamma5}) and (\ref{sigma}) gives
\begin{mathletters}
\begin{eqnarray}
{\cal I}_1({\bar\psi}^{\cal E},\psi^{\cal E})&=&
\frac{f^2}{2}({\bar\psi}^{\cal E}_R\psi^{\cal E}_L)^2,\\
{\cal I}_2({\bar\psi}^{\cal E},\psi^{\cal E})&=&
\frac{f^2}{2}({\bar\psi}^{\cal E}_R\sigma^{\mu\nu}\psi^{\cal E}_L)^2,
\end{eqnarray}
\end{mathletters}
where we have suppressed internal symmetry indices. 

We shall now derive approximate Fredholm integral equations which lead to linear eigenvalue
equations for the mass spectra of quarks and leptons.  By applying the equation of motion 
(\ref{eqsmotion}) to $\psi(x_1)$ in (\ref{taufunctions}) and integrating the resulting equation
using
\begin{equation}
i\gamma_\nu\frac{\partial G(x)}{\partial x^\nu}=\delta^4(x),
\end{equation}
we get
\begin{eqnarray}
\tau(x_1...\vert y_1...)=-\int d^4x'G(x_1-x_2)\langle 0\vert T({\cal N}(x')\psi(x_2)...\nonumber\\
{\bar\psi}(y_1)\vert\psi\rangle+\langle 0\vert T(\psi_0(x_1)\psi(x_2)...{\bar\psi}(y_1)...)
\vert\psi\rangle.
\end{eqnarray}
We introduce
\begin{equation}
S_0(x-y)=\langle 0\vert T(\psi_0(x){\bar\psi}(y)\vert 0\rangle
\end{equation}
and obtain
\begin{eqnarray}
\langle 0\vert T(\psi_0(x_1)\psi(x_2)...{\bar\psi}(y_1)...)\vert\psi\rangle\nonumber\\
=S_0(x_1-y_1)(-1)^m\langle 0\vert T(\psi(x_2)...{\bar\psi}(y_2)...)\vert\psi\rangle\nonumber\\
+S_0(x_1-y_2)(-1)^m\langle 0\vert T(\psi(x_2)...{\bar\psi}(y_1){\bar\psi}(y_3)...\vert\psi\rangle
+...\,.
\end{eqnarray}
From this expression we get
\begin{equation}
S(x-y)=S_0(x-y)-\int d^4x'G(x-x')\langle 0\vert T({\cal N}(x'){\bar\psi}(y))\vert 0\rangle.
\end{equation}

Let us now consider the lowest order $\tau$-function obtained after eliminating all higher
$\phi$-functions:
\begin{equation}
\tau(x)\equiv \phi(x)=\langle 0\vert\psi(x)\rangle.
\end{equation}
We shall assume that the TTD method leads to an
integral equation which produces an eigenvalue equation for the quark and lepton state of the
form
\begin{equation}
i\gamma_\nu\frac{\partial}{\partial x^\nu}\phi(x)=i\gamma_\nu\frac{\partial}{\partial x^\nu}\int
d^4x'Q(x-x')\phi(x').
\end{equation}
Then for the two-point function we expect to get an integral equation :
\begin{equation}
i\gamma_\nu\frac{\partial}{\partial x^\nu}S(x-y)=i\gamma_\nu\frac{\partial}{\partial x^\nu}\int
d^4x'Q(x-x')S(x'-y)+M(x-y).
\end{equation}
In momentum space this gives
\begin{mathletters}
\begin{eqnarray}
{[1-Q(p^2)]}\gamma^\nu p_\nu\phi(p)&=&0,\\
{[1-Q(p^2)]}\gamma^\nu p_\nu S(p)&=&M(p^2),
\end{eqnarray}
\end{mathletters}
and
\begin{equation}
S(p)=\frac{\gamma^\nu p_\nu M(p^2)}{p^2[1-Q(p^2)]}.
\end{equation}

An approximate equation of motion for the quark or lepton state is
\begin{equation}
\gamma_\nu\frac{\partial}{\partial x^\nu}\tau(x)=-\int d^4x'{\rm Tr}[G(x-x'){\cal R}(x'-x)]
\tau(x-x'),
\end{equation}
where the function ${\cal R}$ will be a product of local propagators $S(x-x')$ and regularized
propagators ${\cal S}(x-x')$. In momentum space this becomes
\begin{equation}
(P\cdot\gamma)\tau=\frac{i}{(2\pi)^8}\int \frac{d^4pd^4q{\cal
R}(p,q)L(P,p,q)}{p^2+\kappa^2-i\epsilon}\tau.
\end{equation}
When this expression is analytically continued into Euclidean momentum space, the function 
${\cal R}(p,q)$ will damp out the integrals for large momenta. We now obtain an eigenvalue
equation for the lepton and quark masses:
\begin{equation}
\biggl[1-E\biggl(\lambda_f^2,\frac{\kappa^2}{\Lambda^2_F}\biggr)\biggr]
(P\cdot\gamma)\tau=0,
\end{equation}
where $\lambda_f^2=P^2/\kappa^2$. The discrete poles of the two-point function should then
coincide with the physical quark or lepton masses at the values $P^2_i=m_i^2$.

In practice, it will be necessary to approximate the two-point functions $S(x-x')$ and 
${\cal S}(x-x')$, because in general these functions are complicated solutions of the equations of
motion. An approximation that can be used in calculations is
\begin{mathletters}
\begin{eqnarray}
S(x-x')\approx i(2\pi)^{-4}\int d^4p\frac{\gamma^\nu p_\nu\exp[-ip\cdot(x-x')]}
{p^2+\kappa^2-i\epsilon},\\
{\cal S}(x-x')\approx i(2\pi)^{-4}\int d^4p\frac{\gamma^\nu p_\nu\Pi(p^2)\exp[-ip\cdot(x-x')]}
{p^2+\kappa^2-i\epsilon},
\end{eqnarray}
\end{mathletters}
and the regularizing distribution is chosen to be 
\begin{equation}
{\cal E}(p^2)=\exp(-p^2/2\Lambda_F^2).
\end{equation}

The quarks can form composite bound meson states $\vert M\rangle$ which can be obtained
from two-point $\tau$-functions 
\begin{equation}
\tau(x\vert x)\equiv \phi(x\vert x)=
\langle 0\vert T({\psi}(x)\bar\psi(x))\vert M\rangle.
\end{equation}
The kernel for the eigenvalue equation to lowest approximation will have the form:
\begin{eqnarray}
K(x-x')=-{\rm Tr}[G(x-x'){\cal V}(x'-x)+{\cal V}(x-x')G(x'-x)]\nonumber\\
=(2\pi)^{-8}i\int d^4pd^4q\exp[-i(p-q)\cdot (x'-x)]
\biggl[\frac{{\rm Tr}(\gamma^\mu p_\mu \gamma^\nu q_\nu){\cal V}(q^2)}{p^2}\nonumber\\
+\frac{{\rm Tr}(\gamma^\mu p_\mu\gamma^\nu q_\nu){\cal V}(p^2)}{q^2}\biggr].
\end{eqnarray}
The Fourier tranformation of this kernel is
\begin{eqnarray}
K(P)=\int d^4xK(x-x')\exp[iP\cdot (x-x')]\nonumber\\
=-(2\pi)^{-4}i d^4pp\cdot(P-p)\biggl[\frac{{\cal V}(p^2,\kappa^2)}{(P-p)^2}
+\frac{{\cal V}((P-p)^2,\kappa^2)}{p^2}\biggr].
\end{eqnarray}

The function ${\cal V}$ will damp out the kernel $K$ when it is continued into Euclidean
momentum space. The meson quark-antiquark bound states will be determined by the eigenvalue
equation:
\begin{equation}
1+W\biggl(\lambda_M^2,\frac{\kappa^2}{\Lambda_F^2}\biggr)=0,
\end{equation}
where the eigenvalues will be obtained in terms of the parameter
$\lambda_M^2=P^2/\kappa^2=\kappa_M^2/\kappa^2$.

By using the three point $\tau$-functions an eigenvalue equation can similarly be derived for the
three-quark hadron mass spectrum:
\begin{equation}
\tau(x_1,x_2,x_3)\equiv \phi(x_1,x_2,x_3)=\langle 0\vert T(\psi(x_1)\psi(x_2)\psi(x_3))\vert
0\rangle.
\end{equation}
The confining force for the meson and baryon bound state integral equations should be described
by the regularized QCD gauge interactions, which have to be included with the four-fermion
interactions;
the chirality of the interactions and their symmetry breaking must be taken into account in
calculations of the mass spectrum. 

Finally, from the four-point $\tau$-function
\begin{equation}
\tau(x_1,x_2,x_3,x_4)=\langle 0\vert T(\psi(x_1)\psi(x_2)\psi(x_3)\psi(x_4))\vert\psi\rangle,
\end{equation}
we can calculate the coupling constants of our model, using the TTD approximation in the
neighborhood of the particle poles for a particular choice of the external momenta of the Fourier
transformed four-point function.  Thus, we have provided a scheme for determining the particle
mass spectrum and the coupling constants in our extension of the standard model.

\section{Conclusions}

We have formulated a perturbatively finite, unitary and gauge invariant field theory model of
particles, based
on the product group: $SO(3,1)\otimes SU(3)\otimes SU(2)\otimes U(1)$. All the tree graphs for
the gauge theory are local and
yield the same predictions as the tree graphs in the standard model, including the required Higgs
exchange graphs that guarantee unitarity and renormalizability in the regulated limit 
$\Lambda_F\rightarrow\infty$. The quantum gravity theory is perturbatively finite to all orders,
so it provides a consistent formalism for quantum gravity at least in the perturbative regime. The
graviton tree graphs are purely local and should reproduce
the classical GR theory in the low-energy region together with the standard experimental
agreement of GR. An extended nonlinear gauge symmetry guarantees the decoupling of all
unphysical modes in the theory, using the shadow field formalism and the functional integral
technique together with the BRS ghost formalism, which can be consistently incorporated in the
shadow field formalism.

There is a fundamental energy scale in the theory, namely, the physical scale
$\Lambda_F=1/\sqrt{G_F}\sim 300$ GeV. This constant replaces the
arbitrary, unphysical cutoff in the standard local point particle model, and we showed how 
the Higgs sector hierarchy (`naturalness') problem is resolved, since the quadratic
Higgs mass $\delta m_H^2$ is protected from becoming too large and unstable. For large
$m^2_H/\Lambda_F^2$ the quadratic Higgs mass and the renormalization
coupling constant $\delta\lambda$ for the quartic Higgs interaction are exponentially damped, so
that the Higgs sector is again protected from unstable radiative corrections.  We anticipate that
for the graviton interactions the radiative corrections produced by graviton loops, which grow
faster than $\ln(p^2/\mu^2)$ will also be protected from becoming unstable. This removes one of
the most unsatisfactory features of the standard model, based on local, renormalizable point
interactions for the $SU(3)\otimes SU(2)\otimes U(1)$ gauge and Higgs sector and the local
Lorentz frame $SO(3,1)$ graviton sector. 

Because we have eliminated the Higgs hierarchy problem, there is no
need to introduce supersymmetry into the model. If supersymmetric partners are discovered by
the LHC experiments, then we would of course be required to include them within the FQFT
scheme, which does provide the advantage that it naturally maintains supersymmetry to all
orders of perturbation theory without introducing artificial cutoff and dimensional regularization
schemes.

We have described a self-consistent program to calculate the fermion mass spectrum from
truncated Fredholm integral equations, derived from dynamical equations of motion for the
quark and lepton fields. If this program succeeds in predicting a mass spectrum consistent with
the experimental data, then a considerable number of unknown parameters has been determined
in our development of the standard model using FQFT. In fact, it would then seem less pressing
to search for a ``unified theory" of elementary particles.

If this model can be made to succeed, then we would have achieved a significant economy in the
number of particles, which would correspond to the already discovered bosons and fermions,
with the exception of the Higgs particle. Superstring theory is purported to be the only viable
quantum gravity theory which is both ultraviolet finite and unitary. The theory is intrinsically
nonlocal for both the tree graphs and for the quantum loop corrections. The finite quantum field
theory we have described is only nonlocal at the quantum loop level and it does not require
exotic dimensions, or infinite towers of undesired particles both in the massive and massless
sectors. Also, it provides a self-consistent quantum gravity theory with only one massless particle
-- the graviton. We have formulated the quantum gravity theory at the perturbative level, and we
conjecture that it will be perturbative even at Planck energies, because the FQFT gravity theory
protects leading graviton loop corrections of order $\Lambda^2_F$ from becoming large for
$\Lambda_F\sim 300$ GeV. This is an attractive scenario for quantum gravity, for 
non-perturbative quantum gravity theory is a murky subject at present in field theory as well as in
string theory.

An important aspect of FQFT is that it can avoid the non-perturbative instabilities that beset
string theory and other nonlocal field theories. This strengthens our belief that FQFT can be the
basis of a self-consistent fundamental field theory with the finite scale $\Lambda_F$.

One prediction of our theory  is {\it that the proton is stable.} This currently agrees with the latest
experimental findings. The problem of proton decay is a serious one in unified theories and in
superstring theories. To avoid fast proton decay in these latter theories, it is necessary to invoke
{\it ad hoc} and artificial mechanisms, which postulate undiscovered parity
operations, new unwanted particle sectors in MSSM, or appeal to unproved non-perturbative
arguments  as in superstring theory. Another problem that is avoided in our model is explaining
how supersymmetry is broken spontaneously and at what energy level this occurs. 

Our FQFT scenario predicts that we have reached the `summit' of particle physics. Apart from
the Higgs particle, there will be
no new particles discovered beyond the top quark at $\sim 176$ GeV.  Therefore, if the scenario
is correct, then particle physics can be
considered a `closed' subject, once the Higgs mass is discovered. The experimental
prediction of the `summit` scenario that the proton is stable can be considered as a signature of
the lack of any experimental detection of proton decay.

There is one underlying feature of all current attempts to extend the standard model: the lack of
solid experimental data at energies above the electroweak scale. This hiatus in particle physics
was exacerbated by the cancellation of the supercollider accelerator, because with this accelerator
we could eventually have been guaranteed access to data at energies at or above 10 TeV. Physics
theories ultimately rely on experimental data to last, for only the conclusive agreement of the
predictions of a theory of particle physics and gravity with experiment can guarantee that a
theory will survive and become a worthy paradigm for future physicists. We can only hope that
for the near future the LHC will yield important results that can provide substantive clues as to
the fundamental structure of matter.

\acknowledgments

This work was supported by the Natural Sciences and Engineering Research Council of Canada.

\end{document}